\documentclass[manuscript,screen]{acmart}

\AtBeginDocument{%
  \providecommand\BibTeX{{%
    \normalfont B\kern-0.5em{\scshape i\kern-0.25em b}\kern-0.8em\TeX}}}

\setcopyright{acmcopyright}
\copyrightyear{2020}
\acmYear{2020}

\usepackage{balance}       
\usepackage[]{hyperref}

\def\Snospace~{\S{}} 

\usepackage{color}
\usepackage{booktabs}
\usepackage{textcomp}
\usepackage{csquotes}
\usepackage{multirow}
\usepackage{makecell,rotating}
\usepackage{enumitem}
\usepackage{tabularx}
\usepackage{caption}
\usepackage{array}
\usepackage[ruled]{algorithm2e}
\usepackage{algpseudocode}
\usepackage{xcolor}
\usepackage{xparse}

\usepackage{rotating}

\graphicspath{{figures/}{pictures/}{images/}{./}} 

\usepackage{tikz}
\usepackage{microtype}        
\usepackage{ccicons}          

\usepackage{todonotes}
\usepackage{xspace}
\usepackage[group-separator={,},detect-all=true]{siunitx}

\newcommand{\PO}{\emph{Path Outlines}\xspace}
\newcommand{\SV}{SPARQL-V\xspace}

\newcommand{\md}[1]{\textcolor[rgb]{0, 0, 0}{#1}}

\newcommand{\entity}[1]{\texttt{#1}}

\begin{document}

\title{Path Outlines: Browsing Path-Based Summaries of Knowledge Graphs}






\author{Marie Destandau}
\email{marie.destandau@inria.fr}
\affiliation{
  \institution{Université Paris-Saclay, CNRS, Inria, LRI}
   \city{Orsay}
  \country{France}
}

\author{Olivier Corby}
\email{olivier.corby@inria.fr}
\affiliation{
  \institution{Wimmics, Inria}
  \city{Sofia-Antipolis}
  \country{France}
 }

 \author{Jean Daniel Fekete}
\email{jean-daniel.fekete@inria.fr}
\affiliation{
  \institution{Université Paris-Saclay, CNRS, Inria, LRI}
  \city{Orsay}
  \country{France}
}
 
 \author{Alain Giboin}
\email{alain.giboin@inria.fr}
\affiliation{
  \institution{Wimmics, Inria}
  \city{Sofia-Antipolis}
  \country{France}
 }

\renewcommand{\shortauthors}{Destandau et al.}

\begin{abstract}
Knowledge Graphs have become a ubiquitous technology powering search engines, recommender systems, connected objects, corporate knowledge management and Open Data. They rely on small units of information named triples that can be combined to form higher level statements across datasets following information needs. But data producers face a problem: reconstituting chains of triples has a high cognitive cost, which hinders them from gaining meaningful overviews of their own datasets. We introduce \emph{path outlines}: conceptual objects characterizing sequences of triples with descriptive statistics. We interview 11 data producers to evaluate their interest. We present \PO, a tool to browse path-based summaries, based on coordinated views with 2 novel visualisations. We compare \PO with the current baseline technique in an experiment with 36 participants. We show that it is 3 times faster, leads to better task completion, less errors, that participants prefer it, and find tasks easier with it.
\end{abstract}




 \begin{teaserfigure}
    \frame{\includegraphics[width=\textwidth]{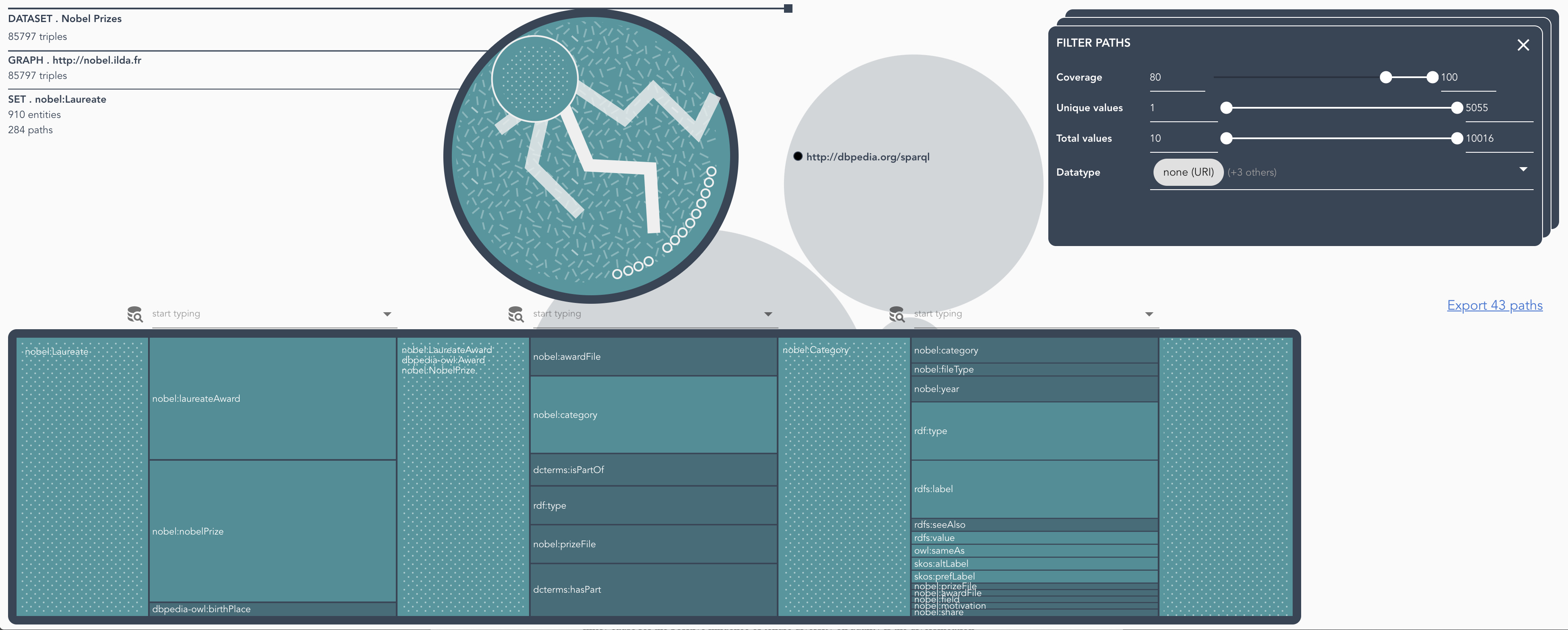}}
    \caption{\PO displays the analysis of paths of depth 3 for the set of \texttt{Laureate}s in the Nobel dataset. The user has used the filter panel to see only the paths describing more than 80\% of the entities: from the initial 80 paths of depth 3, only the 43 paths are left, other are filtered out. The user is currently hovering a property in the second column, which highlights in other columns all properties involved in sequences going through it. Clicking on this property would filter out properties that are not highlighted.}
    \Description{A screenshot of the tool \PO displaying paths of depth 3 for \texttt{Laureate}s in the Nobel dataset. The upper part of the screen is divided between the context zone on the left and the filter panel on the right. The context zone shows the dataset, the set of entities and the length of paths that have been. The lower part of the screen displays the path browser on the left. The right part is left blank for the detail panel, that will appear when a specific path is selected.}
    \label{fig:teaser}
  \end{teaserfigure}

\maketitle

\section{Introduction}
Knowledge Graphs (KG) have become, in recent years, an ubiquitous technology powering search engines~\cite{uyar2015evaluating}, recommender systems~\cite{wang2018ripplenet} and connected objects~\cite{le2016graph}. Institutions~\cite{shadbolt2013linked} and communities rely on them to publish and share their data in an interoperable format~\cite{erxleben2014introducing}.  A growing number of companies use them for Corporate Knowledge Management, to support open innovation and market intelligence strategies, as it is acknowledged that ``the  value [of data] is directly proportional to the interlinkedness of the data''~\cite{pan2017exploiting}.

Knowledge graphs can be composed of one or several data sources, also called Linked Datasets. The interoperability is made possible by the representation of information according to a common framework, the Resource Description Framework (RDF)~\cite{Carroll:04:RDF}.
RDF information is atomized in small units named \textit{triples}. 
The triples can be combined to form complex statements depending on information needs. For instance, a triple in the Nobel Prizes dataset stating that ``Marie Curie is affiliated to Sorbonne University'', and another that ``Sorbonne University is located in Paris'' can be combined into ``Marie Curie is affiliated to Sorbonne University in Paris''. The chaining can be extended to other linked datasets. Such a structure is very expressive and powerful.

However, a drawback of this expressivity is that it makes it difficult for data producers to have an accurate overview of their data and assess
their quality~\cite{troullinou2017ontology}.
\emph{Our goal is to provide meaningful overviews at the right level of abstraction to assess their quality and eventually improve it}.
Important information about an entity is often 2 or 3 triples away from it, and current summary approaches fail to address chains of properties, also named \emph{paths}.
Most existing tools consider only triples, leaving aside all the statements that can be produced by chaining them. Other tools show summary graphs, presented as node-link diagrams, but their labels are difficult to read, and often laid out in various directions, barely allowing to follow paths.
Given the large number of properties even in small databases, the node-link diagrams are either cluttered and unreadable, or reduced to the most frequent classes and properties, offering a very partial overview of the paths available. They also provide metrics, but only
at the triple level, displayed when users select an element in the diagram. In all cases, it is somehow possible to mentally recombine the paths, but this implies a high cognitive load~\cite{destefano2007cognitive}. Furthermore, combining statistics about triples does not provide  statistics about paths. With a summary of the Nobel dataset stating that the database contains 911 laureates, 75\% being affiliated to a university, and 525 Universities, 85\% being located in a city, one could deduce that a laureate can have an affiliation that is located in a city, but there would be no way to know the percentage of laureates that actually do. 

As Marchionini and Shneiderman stated in an early paper about hypertext systems, ``key design issues include finding the correct information unit granularity for particular task domains and users''~\cite{Marchionini1988}. 
We posit that
current RDF summary approaches are limited by their granularity, 
and that the \emph{path} provides a meaningful granularity and expressivity to summarise Knowledge Graphs for data producers. 
We introduce \textit{path outlines}, conceptual objects characterizing sequences of triples with descriptive statistics. To provide an overview, and
allow producers to determine which are of interest to them, we design and implement an interface supporting the Information Seeking Mantra: ``Overview first, zoom and filter, then details-on-demand''~\cite{shneiderman1996eyes}. Based on coordinated views with 2 novel visualisations, it allows to represent a very large number of \textit{path outlines}, browse through them and inspect their metrics. 
Our contributions are: 
\begin{itemize}
\item the concept of \emph{path-based summaries} and an API to analyse them;
\item \PO, the design and implementation of an open-source tool to browse path-based summaries; and
\item a controlled experiment to evaluate the tool against Virtuoso SPARQL query editor as a baseline.
\end{itemize}

First, we introduce the difficulties to represent RDF data and visualise paths, and we discuss related work. Then we define the concept of \emph{path outlines} to support path-based summaries, and we describe our API to analyse them. Next, we report on the interview of 11 data producers to evaluate their understanding and interest. After that, we present \PO, our tool based on coordinated views representing the features of \emph{path outlines}, to browse such summaries. Eventually, we conduct a use-case based evaluation of \PO with 36 participants, in which we compare it with the Virtuoso SPARQL query editor as baseline. 

%
\section{Background and Related Work}
We discuss the difficulties to represent RDF data and paths, the types of summaries which are currently available, and the difficulty of writing and running queries for path-based summary information.

\subsection{RDF Data}\label{sec:concepts}
\begin{figure}[h!]
	\includegraphics[width=\columnwidth]{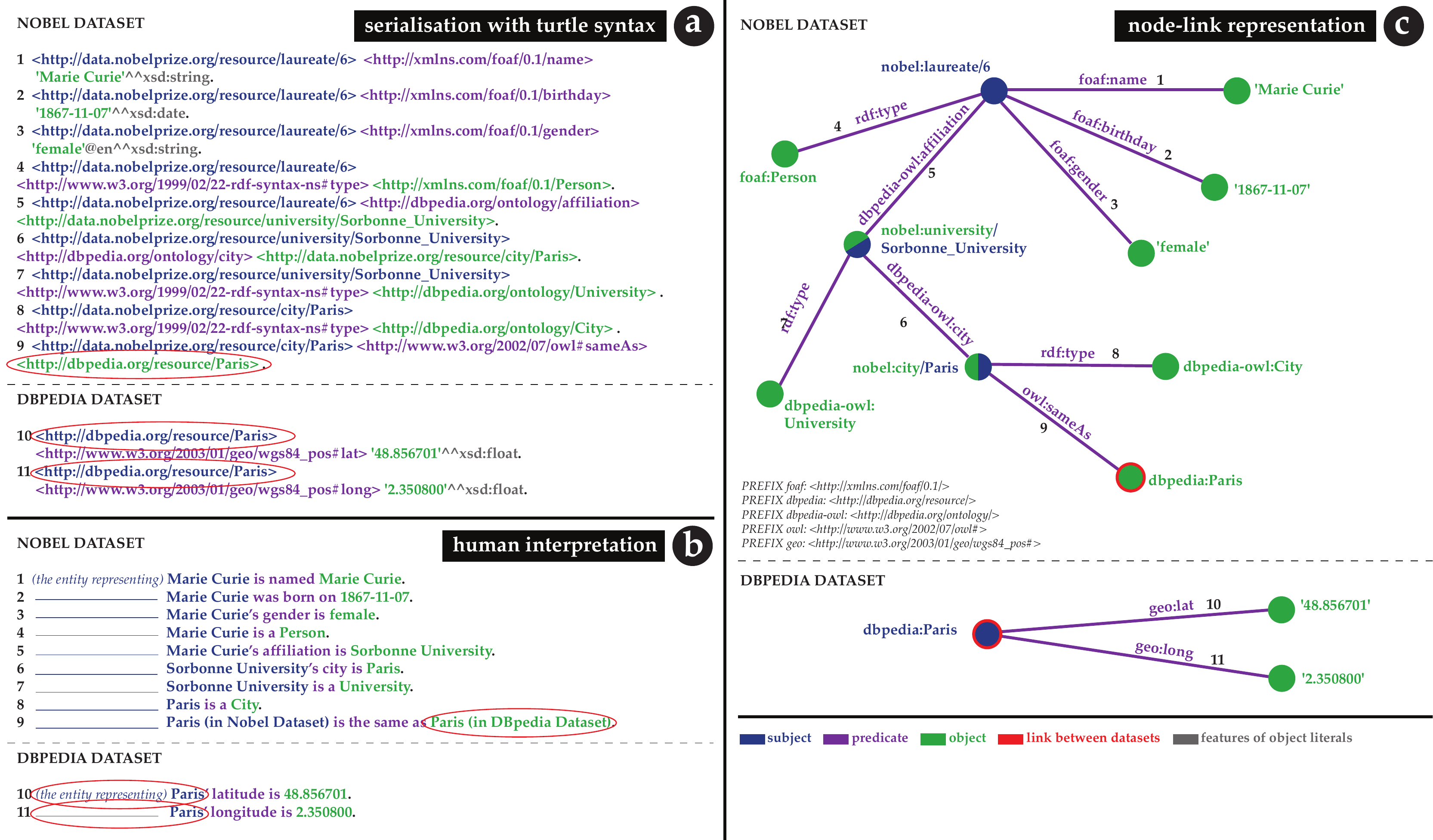}
        \caption{Samples extracted from Nobel (9 triples) and DBpedia datasets (2 triples) comparing 3 representations: a) a serialisation with Turtle syntax, which is meant for machines, but verbose and difficult to decipher for humans; b) an interpretation of each triple as a sentence understandable by humans to help understand the chaining mechanism; c) a visualisation as a node link diagram showing how triples are connected and can be chained; it is easy to follow the paths but difficult to read their labels and make sense of them. The samples are interconnected; it is possible to combine them. Full datasets contain respectively \num{87422} and \num{185404534} triples (on 2019-09-07).}
    	\Description{The figure compares 3 representations for RDF samples: a) the Turtle serialisation shows the raw triples, made of URIs and literals; it is verbose and meant for machines, difficult to decipher for humans; b) the interpretation of each triple as a sentence understandable by humans helps understanding the chaining mechanism; c) the visualisation as node-link diagram shows how triples are connected and can be chained. It also helps understanding but demands efforts to decipher. }
	\label{fig:explain}
\end{figure}

RDF data are collections of statements named triples. Triples are composed of a \emph{subject}, a \emph{predicate} and an \emph{object}, as shown in \autoref{fig:explain}. Subjects and predicates are always \emph{Uniform Resource Identifier (URIs)}. Objects can be URIs ($\ell$.\ 4--9) or literals (e.g., strings, numbers, dates, $\ell$.\ 1--3). The same URI can be the subject and object of several triples ($\ell$.\ 5, 6, and 7 or $\ell$.\ 6, 8 and 9). The triples form a network. The predicate \texttt{rdf:type} ($\ell$.\ 4, 7 and 8) indicates that an entity belongs to a class of resources. Predicates and classes of resources are defined in data models called \emph{ontologies}---somehow equivalent to schemas in SQL databases. For instance, the predicates of the 3 first triples, and the object of the 4th, belong to the FOAF~\cite{Brickley:14} (friend of a friend) ontology, dedicated to the description of people and their relationships. 
Literals can be typed, and string literals can be associated with a language (\autoref{fig:explain}-a, grey colour). 
URIs can be prefixed for better readability, as in \autoref{fig:explain}-c: the beginning, common to several URIs, is given a prefix (a short name), e.g. \texttt{foaf:} instead of \texttt{http://xmlns.com/foaf/0.1/}.
Formally, a RDF graph is a set of triples $t = (s, p, o)$, with $s \in U$,  $p \in U$ and  $o \in U\cup L$. $U$ is the set of URIs, and $L$ the set of literals in the graph.

RDF data are interlinked: a dataset can reference an entity produced in another one (red colour). When this happens, a chain of statements can jump from one dataset to another: the triples in Nobel Dataset \textit{la Sorbonne is in Paris, Paris entity in Nobel is equivalent to Paris entity in DBpedia} can be completed by those from DBpedia: \textit{Paris' latitude is 48.856701, Paris' longitude is 2.350800}. They can be queried jointly through \emph{federated} queries.

\subsection{Visualisation of RDF Data}
Node-link diagrams (\autoref{fig:explain}-c) are often used to represent RDF datasets~\cite{po2020linked}. They accurately render their structure and are theoretically appropriate to easily recombine paths~\cite{novick2006understanding}. 
However, the readability of paths as sequences is very limited. Huang and Eades remark that people try to read paths from left to right and top to bottom, even when the layout and task require another direction~\cite {huang2005people}. Van Amelsvoort et al.\ demonstrate that reading behaviours were influenced by the direction of elements~\cite{van2013importance}. Ware et al.\ show that good continuity, edge crossing and path length influence the effectiveness of visually following a path~\cite{ware2002cognitive}. A specific type of node-link diagrams, node-link trees, seem to be more efficient for tasks related to following paths, traversing graphs~\cite{novick2006understanding, novick2001matrix}, and reading paths~\cite{lee2006treeplus}, probably because they constrain the flow in one direction. In their survey on the readability of hypertext, DeStefano and Lefevre mention several studies showing that the multiplication of possibilities impacts readability negatively~\cite{destefano2007cognitive}, supporting the same idea. In contrast, PathFinder~\cite{partl2016pathfinder} lays flat all possible paths for the graph, allowing to read them easily, but this results in a very long list needing to be paginated even when the graph is small. 
For a data producer, gaining an overview of the paths in her own dataset is an unresolved problem. There is a need for a layout that would allow to preserve
their readability as sequences of statements to let users make sense
of them while allowing to see all of them at a glance, easily filtering from the overview to the detail, and selecting them; this is what our tool does.


\subsection{RDF Summaries for Data Curation}
A RDF summary is a concise description of the content of a dataset, sometimes characterised by descriptive statistics. We consider summaries that are meant for data producers, with the purpose of giving an overview of a dataset. 
Data profiling systems are tools presenting statistics about the data, such as LODStats~\cite{ermilov2013}, ProLOD~\cite{Bohm_2010}, LOUPE~\cite{mihindukulasooriya2015loupe} or AETHER~\cite{Makela_2014}. They typically present measures of atomic elements. While such summaries are complete and accurate, they give little information about the content. For a data producer, knowing that, for instance, 37\% of all entities have a \texttt{rdfs:label} does not indicate what those labels are about, and is not very helpful to find missing information. Information becomes more meaningful with more context, like considering properties relatively to \textit{subjects} with a specific \texttt{type}~\cite{issa2019revealing} (the number of \texttt{foaf:Document} having a \texttt{rdfs:label}), or to \textit{objects} with a specific \texttt{type}~\cite{spahiu2016abstat, dudavs2015dataset, dudavs2015discovering} (the number of \texttt{Persons} having a \texttt{birthplace} that is a \texttt{City}). This leads to more interpretable summaries, but is still limited to triples, not considering chains of statements.

Another approach consists in reconstituting a representative graph as the summary. Smallest representative graphs are used for machine consumption, but are too big to be presented to users. User-oriented summaries limit summary graphs to the most represented classes, and the most represented direct properties between them~\cite{troullinou2017ontology, Troullinou2018, weise2016ld}, which make them graspable, yet very incomplete. Those summary graphs preserve access to chains of statements, but the statistics are produced and accessible at the triple level only, when users select an edge on the diagram. Furthermore, as we explained in the previous subsection, the readability of node-link diagrams is limited.
Our approach considers an intermediate level as a unit for summaries: the path. This allows us to summarise statements at a granularity that better matches data producers' needs, including the possible extensions of a path in interlinked datasets, and to provide metrics at this level of granularity.


\subsection{Querying Summary Information}
SPARQL, the main query language for RDF data, provides a syntax to query triples and paths in a graph. 
SPARQL also provides aggregation operators that can be applied to the elements we mentioned: entities, properties, triple patterns, possibly specifying the type of the subject and / or of the object, and deeper path patterns. However, queries combining aggregation and paths patterns are complex, and complex queries raise both technical and conceptual issues, as reported by Warren et al~\cite{warren2018using}. From a technical point of view, the cost of a query increases with the number of entities and the length of paths to evaluate. It is also impacted by the fact that a query is federated (targets several datasets), often resulting in network and server timeouts and errors that are difficult to manage in existing systems.
From a conceptual point of view, summarising paths patterns for a large number of entities is not a simple mental operation. The task can be alleviated by tools to assist writing queries. YASGUI~\cite{rietveld2017yasgui} offers  auto-completion, syntax colouring and prefix handling. SPARKLIS~~\cite{ferre2017sparklis} offers the possibility of discovering the model iteratively, enabling at each step to browse the available possibilities for extending the current path. However, such tools support only part of the task. They can be combined, which requires switching from one to another, and planning and thinking with them remains complicated and error-prone. 
Altogether, there is a need for a tool to facilitate data curation by summarising and visualising paths in RDF data, including extensions to other datasets.


\section{Path Outlines}
To provide metrics at the granularity needed by data producers to make sense of their datasets, we use \emph{paths} and formalise the concept of \emph{path outlines}, making them first-class citizens that can be described, searched, browsed, and inspected.

\subsection{Definition}
\begin{figure}[h]
	\includegraphics[width=0.9\columnwidth]{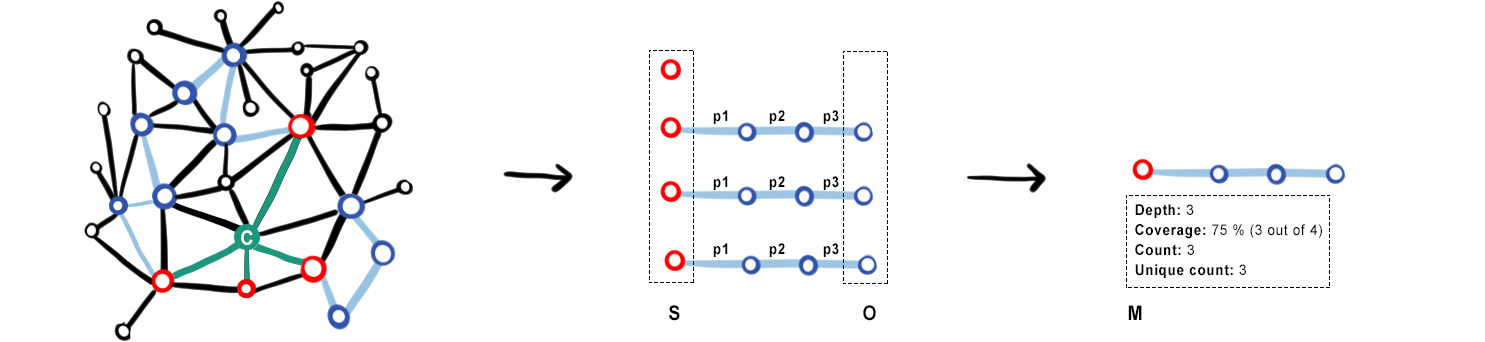}
	\caption{A \emph{path outline}: for a set of entities $S$ (red nodes) sharing a similarity criteria $C$ (green node), a given sequence of properties $p_1/p_2/\ldots/p_n$ (light blue edges) leads to a set of objects $O$. $S$ and $O$ are characterised with a set of measures $M$. One can see that the starting entity for which the path is missing is taken into account in the summary.}
	\Description{Schematic representation of an example of path outline. On the left a node link diagram symbolises the original dataset. The nodes and edges composing the path of interest are colored; it is difficult to untangle them. In the center the paths are extracted and laid flat one under another. One can see that the starting entity for which the path is missing is also taken into account. On the right the summary shows the sequence of triples that is summarised only once, together with the measures that describe it.}
	\label{fig:figurepaths}
\end{figure}

A \emph{path outline} is a conceptual object providing descriptive statistics about a sequence of statements relative to a set of entities. It consists in: 1) a set of entities sharing a similarity criteria (e.g., all the entities of class \texttt{Person}), for which at least one entity is the subject of a given sequence of properties, 2) the sequence of properties, 3) the set of objects at the end of this sequence, and 4) the set of measures relative to the entities and the objects, as schematised in \autoref{fig:figurepaths}.

A path of depth $n$ is a sequence of $n$ triples such that $(s, p_1, o_1), (o_1, p_2, o_2)/\ldots/(o_{(n-1)}, p_n, o)$. Using the SPARQL property path syntax, this could be shortened as $(s, p_1/p_2/\ldots/p_n, o)$. 
To analyse a \emph{path outline}, we start from a given set of entities $S$ sharing a similarity criterion $\texttt{c} \in U$, and we consider a given sequence of properties, such that 
$ \forall s \in S, (s \texttt{ rdf:type c})\,  \exists s \in S, \exists o \in O, (s, p_1/p_2/\ldots/p_n, o) $. $O$ is the set of objects $o$ at the end of the \emph{path outline}. We compute a set of measures $M$ relative to $S$ and $O$, as described in \autoref{tab:table2}. Each measure can be a literal value (e.g., a count), a distribution of values (e.g., the number of unique values for URIs), or a range for numerical values.
\begin{table}[t]
	\small
	\begin{tabular}{ r|>{\raggedright\arraybackslash}m{12cm} } 
		{\bf Measure} & {\bf Description} \\ 
		\hline
		\emph{depth} & number of statements between the set of entities $S$ and the set of objects $O$ \\ 
		\hline
		\emph{coverage} & percentage of entities in the set $E$ for which this path exists \\ 
		\hline
		\emph{count} & total number of objects in $O$\\ 
		\hline
		\emph{unique count} &  number of unique values or URIs for the objects in $O$\\ 
		\hline
		\emph{datatypes} & \makecell[l]{only for literals: data type(s) of objects $O$ at the end of the path}\\
		\hline
		\emph{languages} &\makecell[l]{only for string literals, if specified: list of languages of the objects $O$}\\
		\hline
		\emph{min / max} &\makecell[l]{if numerical values: minimum and maximum value; 
			if strings: first and last value, in alphabetical order} \\
		\hline
	\end{tabular}	
	\caption{Measures characterising a \emph{path outline}}
	\label{tab:table2}
\end{table}

\begin{figure}[h]
	\includegraphics[width=\columnwidth]{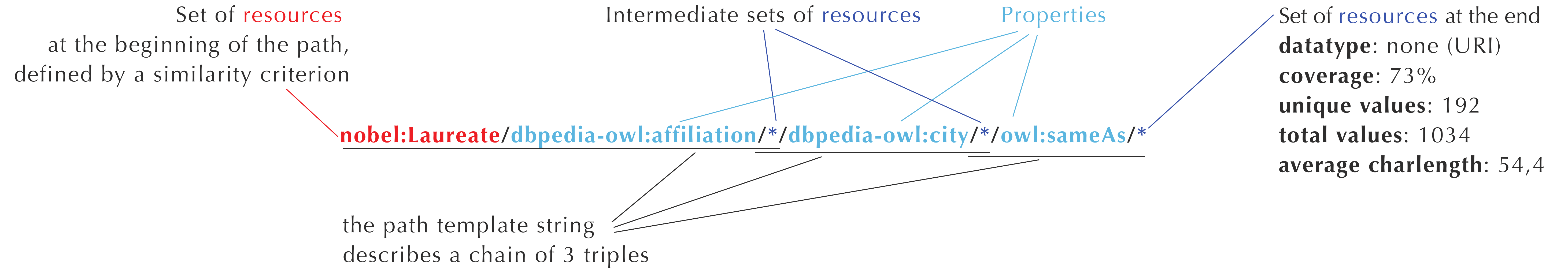}
	\caption{Template string for a \emph{path outline}, summarising the Nobel laureates having an \emph{affiliation}, \emph{located in a city}, \emph{having a similarity link} to another resource. Intermediate sets of resources are designated by stars, indicating that they can be of any type. Those resources are both the objects of the preceding predicate, and the subjects of the next. }
	\Description{The parts composing the template string are separated by slashes. The first part describes the set of resources at the beginning of the path, here the nobel:Laureates, followed by a slash. Then comes the first predicate (dbpedia-owl:affiliation) followed by a slash, followed by a star, indicating that the objects of this predicate can be of any type. The star also represent the subjects of the next triple, it is not repeated. Then comes the next predicate (dbpedia-owl:city), also followed by a star, and the last (owl:sameAs), followed by a star. The last star corresponds to the set of objects O at the end of the path.}
	\label{fig:figureexample}
\end{figure}

To write a path, we defined a syntax inspired from XPath~\cite{clark1999xml} (\autoref{fig:figureexample}). The template string is similar to an XPath query selector: it is a pointer to designate the chains of triples corresponding to the query and summarised by a \emph{path outline}. The syntax is easy to parse at a glance: the elements are separated by a slash (reminiscent of the syntax of file paths in operating systems). The first chunk is the similarity criteria, the number of stars indicates the depth of the path, and they create a visual articulation to separate the other chunks, corresponding to the properties forming the path. It already forms a graphical object revealing the articulations that will show in our visualisations.


\subsection{LDPath API}

To analyse the paths, we developed a specific extension to a semantic framework for Knowledge Graph querying\footnote{reference to a paper, anonymised for submission}. Given an input query, it discovers and navigates paths in a SPARQL endpoint by completing the input query with predicates that exist in the endpoint. LDPath first computes the list of possible predicates and then, for each predicate, counts the number of paths. This is done recursively for each predicate until a maximum path length is reached. The values at the end of each path are analysed to retrieve the features listed in \autoref{tab:table2}.
LDPath can also, for each path, count the number of joins of this path in another endpoint, and compute the list of possible predicates to extend the path by one statement. The values at the end of the extension are also analysed. The software package consists in recursively rewriting and executing SPARQL queries with appropriate service clauses. The API of this extension is made available for other purposes and can be queried independently of \PO\footnote{link to the API, anonymised for submission}. 


\section{User study 1: Validating the Approach}
One of the authors has several years of experience in the Knowledge Graphs community. Taking inspiration from her experience and situations she had observed in professional semantic web meetups and conferences, we designed 6 real-world task scenario involving finding or browsing path-based metrics. For instance, the second scenario was: ``Find the datatypes of the set of values at the end of a path. For example, identify if at the end of certain properties there are alternatively dates or URIs, or check if the date formats typed as such and valid''. We interviewed 11 data producers to validate
our approach. 

\subsection{Participants}
We conducted a fifteen to thirty-minute interview with 11 RDF data producers recruited via email calls on Semantic Web mailing lists and Twitter. Participants belonged to industry~(4), academia~(4) and public institutions~(3). The datasets they usually manipulated contained data from various domains, ranging from biological pathways to cultural heritage through household appliances. All participation was voluntary and without compensation. 

\subsection{Set up and Procedure}
The interview was supervised online through a videoconference system.
We presented each task scenario. We asked participants if they did already perform similar tasks; and if so, how often and by which means; if not, for what reason. We asked them if they would be interested in a tool supporting such tasks. Eventually, we asked them if they could think of other similar or related tasks that would be useful for them. 

\subsection{Results}
We collected answers in a spreadsheet and analysed them with R. 
\begin{figure}[!ht]
	\includegraphics[width=0.45\columnwidth]{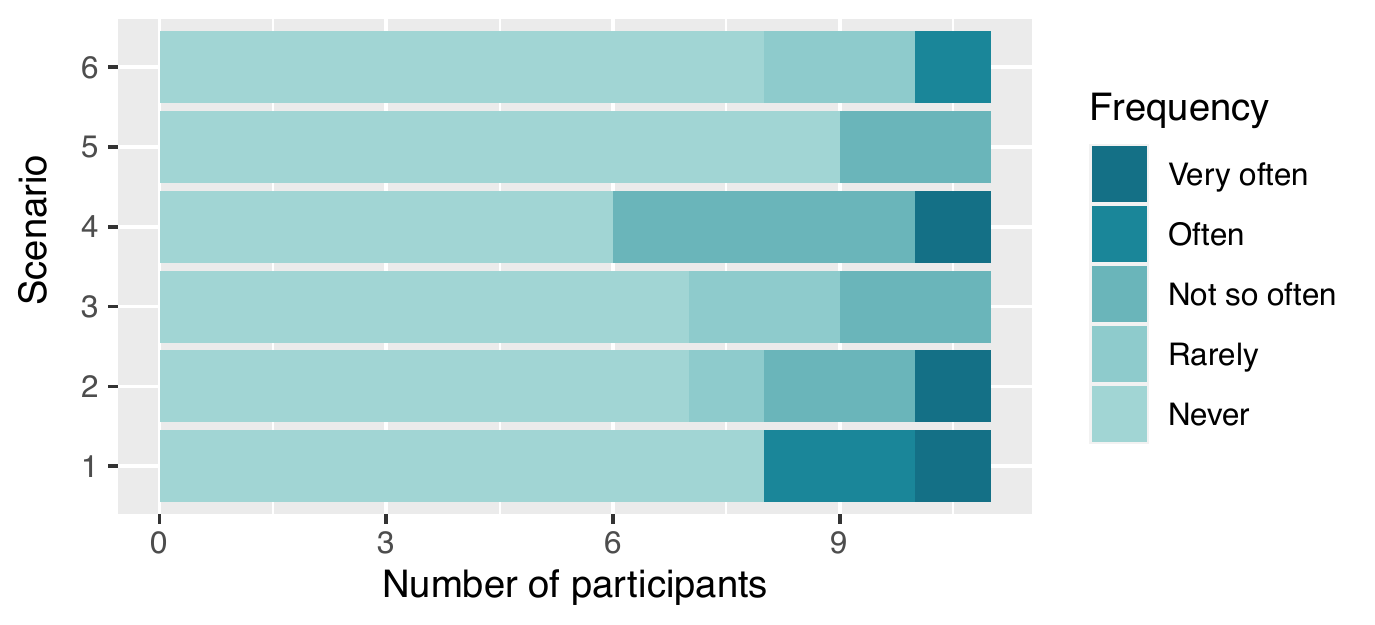}
	\includegraphics[width=0.45\columnwidth]{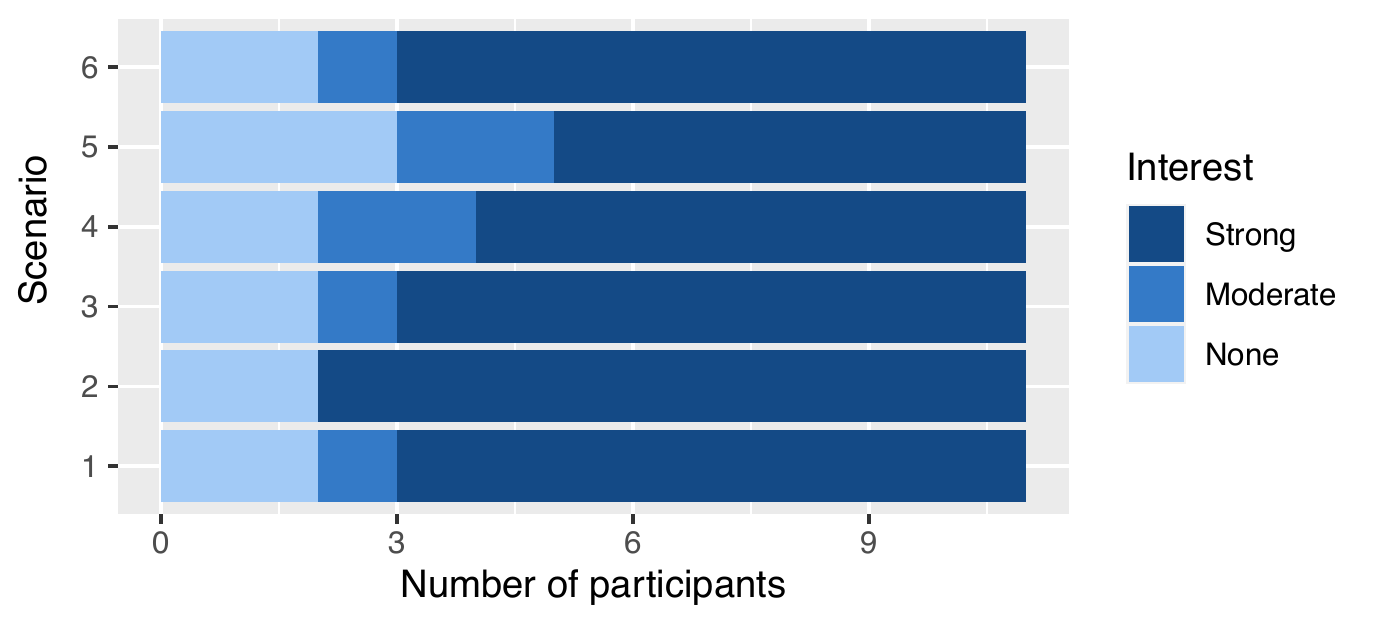}
	\caption{Usage and interest of data producers regarding the scenarios: a) they hardly ever perform similar tasks, b) but would be very interested in a tool supporting them.}
	\Description{On the left, the stacked chart labeled 'a' reports the frequency to which participants perform tasks similar to the ones involved in the scenarios we presented. It can be converted into the following table:
	Scenario         Frequency          Number of participants
	Scenario1       Never                 8
	Scenario1       Often                 2
	Scenario1       Very often          1
	Scenario2       Never                 7
	Scenario2       Rarely                1
	Scenario2       Not so often       2
	Scenario2       Very often          1
	Scenario3       Never                 7
	Scenario3       Rarely                2
	Scenario3       Not so often       2
	Scenario4       Never                 6
	Scenario4       Not so often       4
	Scenario4       Very often          1 
	Scenario5       Never                 9
	Scenario5       Not so often       11
	Scenario6       Never                 8
	Scenario6       Rarely                2
	Scenario6       Often                 1    
	On the right, the stacked chart labeled 'b' reports the interest of participants for the scenarios. It can be converted into the following table:
	Scenario         Degree of interest      Number of participant
	Scenario1       None                          2
	Scenario1       Moderate                    1
	Scenario1       Strong                        8
	Scenario2       None                          2
	Scenario2       Strong                        9
	Scenario3       None                          2
	Scenario3       Moderate                    1
	Scenario3       Strong                        8
	Scenario4       None                          2
	Scenario4       Moderate                    2
	Scenario4       Strong                        7
	Scenario5       None                          3
	Scenario5       Moderate                    2
	Scenario5       Strong                        6
	Scenario6       None                          2
	Scenario6       Moderate                    1
	Scenario6       Strong                        8   
	 }
	\label{fig:figure0}
\end{figure}
\subsubsection{Current Usage of Path-based Metrics} 
A few participants already performed tasks that were similar to the ones in our scenarios, as reported in \autoref{fig:figure0}. 
They used SPARQL query editors~(16)\footnote{the counts in this paragraph correspond to the number of scenarios, not to the number of participants} or  \emph{content negotiation} in the browser~(3): they pasted a URI in the browser to see the triples describing it, and copy-pasted other URIs to continue the chaining, entity by entity. The main reason given for not performing a task or performing it too rarely was \textit{no tool}~(14). Those tasks are actually possible with SPARQL, but participants either did not know how to write the queries or regarded it as so complicated that they would not even consider it as an option. The second main reason was time concerns~(13): the task was regarded as doable, but it would have taken too long to write such queries.

\subsubsection{Interest for Path-based Summaries} 
Two participants had difficulties in relating to the scenarios. Their use of RDF data was focused on querying single entities rather than sets. They did not feel the need for an overview (although one changed his mind, as explained in \autoref{sec:other}).
 Most other participants declared a strong interest (\autoref{fig:figure0}): 3 had already well identified their needs, and the others sounded
 really
enthusiastic that we were able to formulate them. 
Six participants spontaneously mentioned clearly seeing the interest of a tool supporting similar tasks for data reusers, in a discovery context. 
Only one participant suggested a related task: \textit{identify outliers in values of paths typed as numerical values}, involving more advanced metrics on paths than the one we had mentioned.  

\medskip
This interview confirmed the interest of data producers for path-based summaries, and the fact that for those who were already gathering very similar information, a SPARQL query editor was the baseline.

\section{Path Outlines}
To let users browse \emph{path outlines}, we designed an interface based on coordinated views with two new visualisations (\autoref{fig:teaser}). We present the design requirements and the design rationales for the interface, followed by 2 scenarios of use.

\subsection{Design requirements: from overview to detail}
The process of browsing through an information space can be well described by the Information Seeking Mantra: ``Overview first, zoom and filter, then details-on-demand''~\cite{shneiderman1996eyes}: 
The tasks involved in this navigation paradigm are: `find the number of items', `see items having certain attributes', and `see an item with all its attributes'.
The overview is meant to provide context to users, to `gain an overview of the entire collection'. There are several levels of contexts for paths: the dataset, and the starting set of entities. It shall also give them an idea of the main features of the items in the collection, which will allow them to determine what is of interest and what is not, and to progress through the collection, `zoom in on items of interest and filter out uninteresting items', and finally `select an item or group and get details when needed'. The particular difficulty with \emph{path outlines} is that their features are both metrics related to them and the sequences of properties composing them. To address this specificity, our interface combines several coordinated views.

\begin{figure}[!ht]
\includegraphics[width=0.23\textwidth]{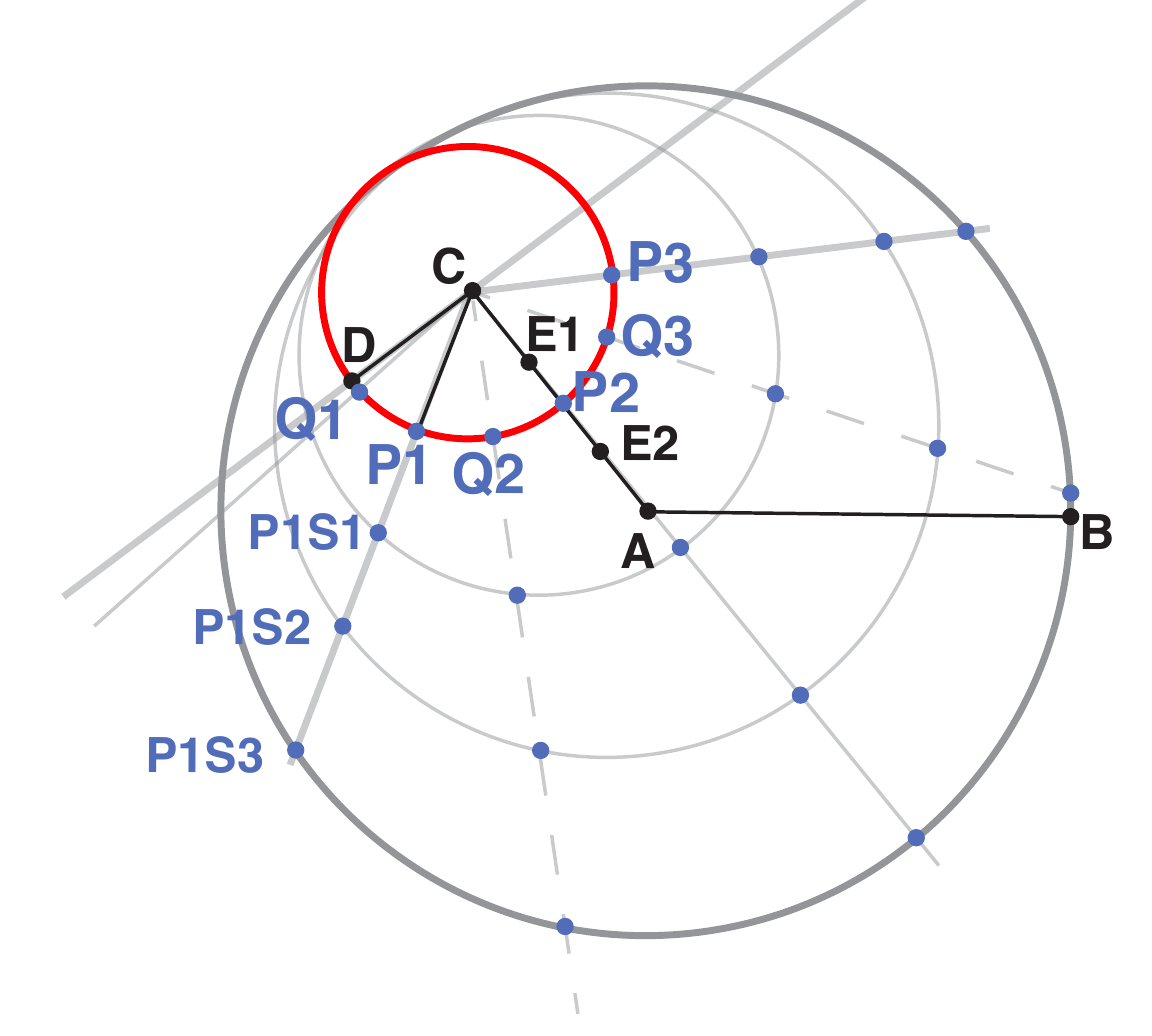}
	\includegraphics[width=0.69\textwidth]{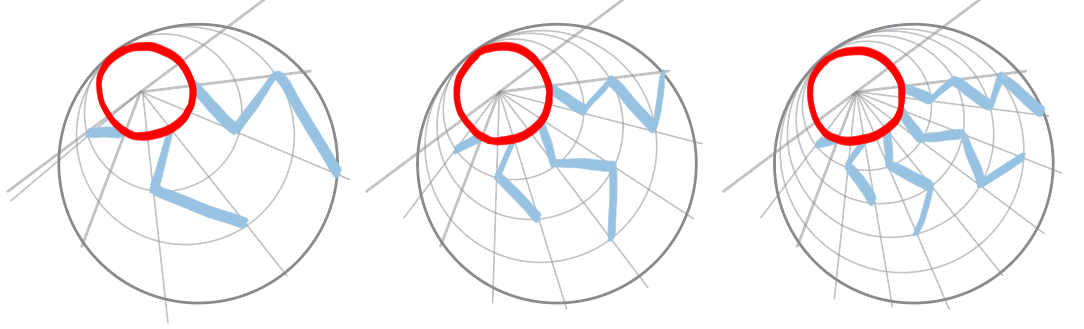}
	\includegraphics[width=\textwidth]{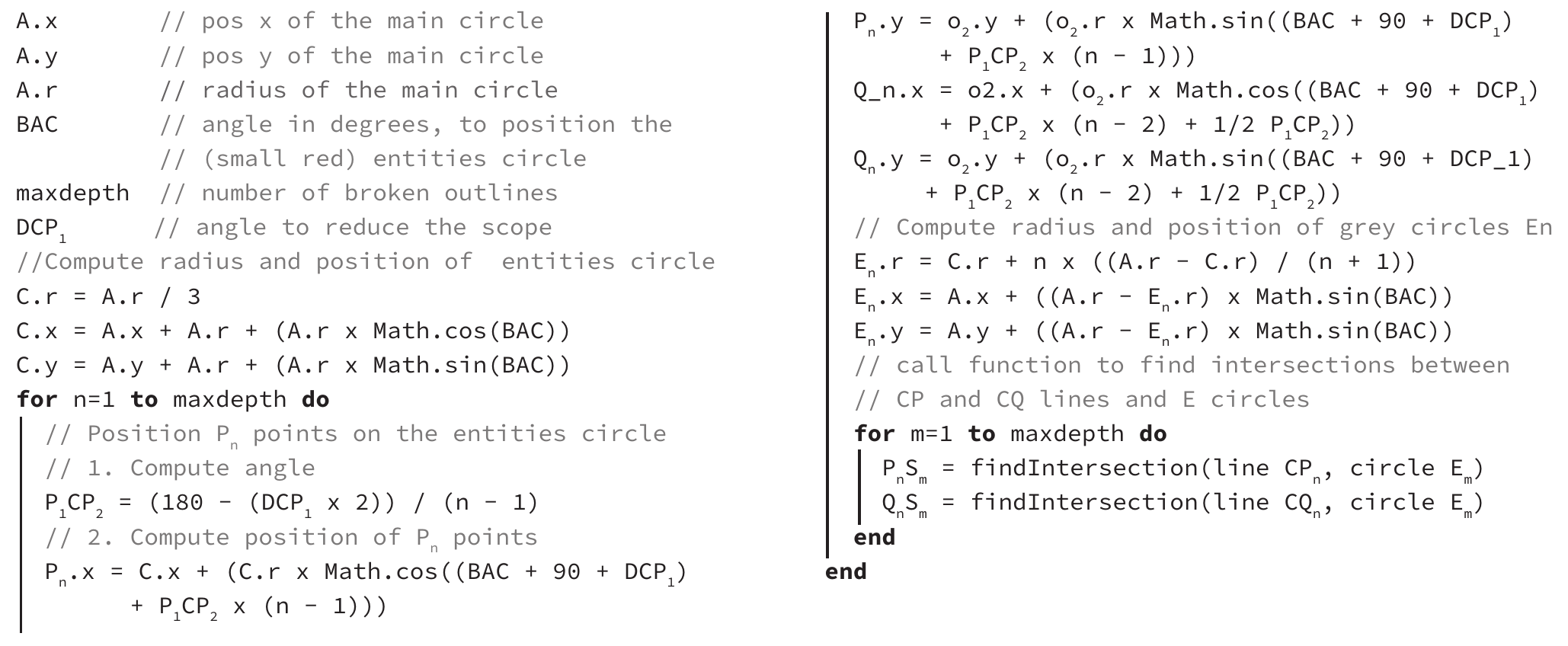}
	\caption{Broken (out)lines algorithm: broken (out)lines are drawn and positioned according to the maximum depth of \emph{path outline}, using geometrical principles to fit in the circle.}
	\Description{Schema showing the lines used to lay out the broken lines. The left circle shows the points that will be used in the pseudo-code. A is the center of the container circle. C is the center of the circle representing the sets of entities, contained in the main circle, its border is near to the border of the container. The broken lines are little 'legs' starting from the small circle and spreading like a fan to take the space available in the container. The pseudo code to draw the lines is the following:
	A.x is the x coordinate of the main circle,
	A.y is the y coordinate of the main circle,
	A.r is the radius of the main circle,
	BAC is the angle in degrees, to position the (small red) entities circle,
	maxdepth is the number of broken outlines,
	DCP1 is the angle to reduce the scope.
	
	//Compute radius and position of  entities circle
	C.r = A.r / 3
	C.x = A.x + A.r + (A.r x Math.cos(BAC)) 
	C.y = A.y + A.r + (A.r x Math.sin(BAC))
	for n=1 to maxdepth do
  		// Position Pn points on the entities circle
  		// 1. Compute angle
  		P1CP2 = (180 - (DCP1 x 2)) / (n - 1)
  		// 2. Compute position of Pn points
  		Pn.x = C.x + (C.r x Math.cos((BAC + 90 + DCP1) + P1CP2 x (n - 1)))
  		Pn.y = o2.y + (o2.r x Math.sin((BAC + 90 + DCP1) + P1CP2 x (n - 1)))
  		Q_n.x = o2.x + (o2.r x Math.cos((BAC + 90 + DCP1) + P1CP2 x (n - 2) + 1/2 P1CP2))
  		Qn.y = o2.y + (o2.r x Math.sin((BAC + 90 + DCP_1) + P1CP2 x (n - 2) + 1/2 P1CP2)) 
  		// Compute radius and position of grey circles En
  		En.r = C.r + n x ((A.r - C.r) / (n + 1))
  		En.x = A.x + ((A.r - En.r) x Math.sin(BAC))
  		En.y = A.y + ((A.r - En.r) x Math.sin(BAC))
  		// call function to find intersections between 
  		// CP and CQ lines and E circles
  		for m=1 to maxdepth do
    		PnSm = findIntersection(line CPn, circle Em)
    		QnSm = findIntersection(line CQn, circle Em)
	 	end
	end	
	
	}
	\label{fig:brokenlines}
\end{figure}

\subsection{Interface: coordinated views to display complex objects}
The interface relies on two new visualisations: the \emph{broken (out)lines} algorithm -- extending a circle packing layout, and the \emph{path browser}. They are coordinated with several filter panels.


\subsubsection{Context overview: circle packing and broken (out)lines}
As users open \PO, they see several datasets laid out with a circle packing algorithm~\cite{COLLINS2003233}. Their size is mapped to the number of triples they contain  (\autoref{fig:steps1}-1). 
Using the filter panel (\autoref{fig:steps1}-2), they can select a specific size range or search by name. When they open it in the foreground (\autoref{fig:steps1}-3), datasets that are linked to it also come to the foreground,  as small bullets laid out on the side (\autoref{fig:steps1}-8). The different sets of entities sharing the same \texttt{rdf:type} in the main dataset are laid out inside in another circle packing, their size corresponding to the number of entities (\autoref{fig:steps1}-4). The filter panel allows to filter by size and name (\autoref{fig:steps1}-6). As they click on one to open it,
other sets become smaller and are aligned on the side to be easily available (\autoref{fig:steps1}-8). The available \emph{path outlines} depths (\autoref{fig:steps1}-7) are laid out with the broken (out)lines algorithm. It relies on simple geometrical principles. The algorithm is described in \autoref{fig:brokenlines}. It is inspired by systems that present an overview of a graph with different possible \textit{cuts} in it, that can be inspected in a coordinated view~\cite{abello2006ask, archambault2010tugging}. The shape of \emph{broken (out)lines} is reminiscent of a node-link diagram so that users can relate it to a representation they already know, and understand what is displayed below in the \emph{path browser} (\autoref{fig:steps1}-9). Associated with the circle, they form a glyph~\cite{Fuchs2017}, a simple symbol meant to be readable, and yet encoding important attributes of the data. By default, \emph{path outlines} of depth 1 are selected. 

\subsubsection{Zoom and filter: the {\em path browser} and filter panel}
\begin{figure}[ht]
	\centering
	\includegraphics[width=\columnwidth]{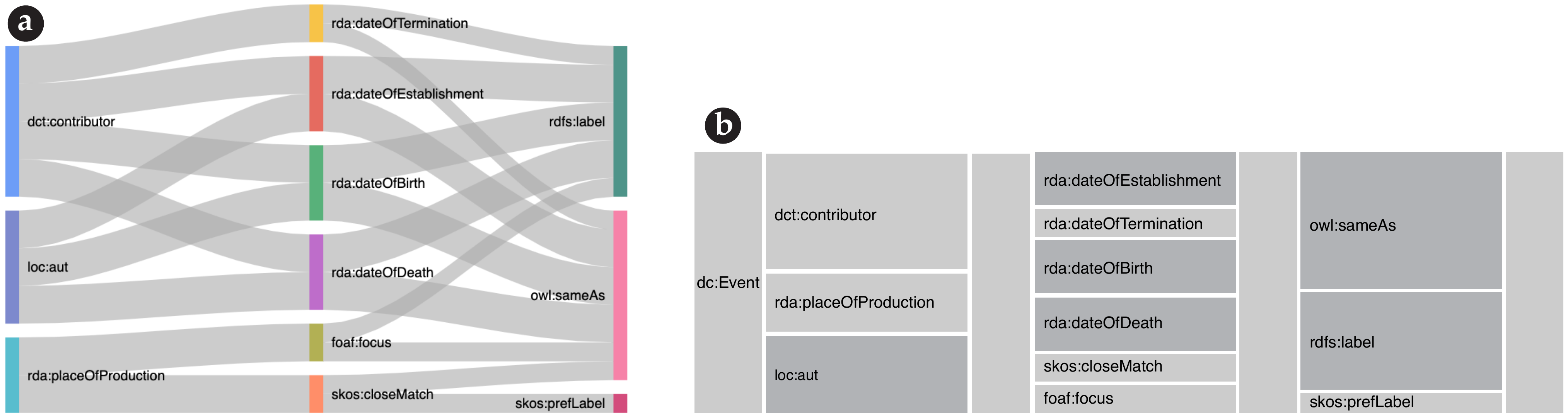}
	\caption{The same 18 \emph{path outlines} displayed in a Sankey Diagram (a) and
          the \emph{path browser} (b). Hovering the property \texttt{loc:aut} highlights all matching sequences. }
     \Description{The Sankey diagram and path browsers both lay out the following paths outlines, showing the combination of sequences:     
dc:Event/dct:contributor/*/rda:dateOfEstablishment/*/rdfs:label/*
dc:Event/dct:contributor/*/rda:dateOfTermination/*/rdfs:label/*
dc:Event/dct:contributor/*/rda:dateOfBirth/*/rdfs:label/*
dc:Event/dct:contributor/*/rda:dateOfDeath/*/rdfs:label/*
dc:Event/dct:contributor/*/rda:dateOfEstablishment/*/owl:sameAs/*
dc:Event/dct:contributor/*/rda:dateOfTermination/*/owl:sameAs/*
dc:Event/dct:contributor/*/rda:dateOfBirth/*/owl:sameAs/*
dc:Event/dct:contributor/*/rda:dateOfDeath/*/owl:sameAs/*
dc:Event/rda:placeOfProduction/*/skos:closeMatch/*/skos:prefLabel/*
dc:Event/rda:placeOfProduction/*/skos:closeMatch/*/owl:sameAs/*
dc:Event/rda:placeOfProduction/*/foaf:focus/*/rdfs:label/*
dc:Event/rda:placeOfProduction/*/foaf:focus/*/owl:sameAs/*
dc:Event/Loc:aut/*/rda:dateOfEstablishment/*/rdfs:label/*
dc:Event/Loc:aut/*/rda:dateOfBirth/*/rdfs:label/*
dc:Event/Loc:aut/*/rda:dateOfDeath/*/rdfs:label/*
dc:Event/Loc:aut/*/rda:dateOfEstablishment/*/owl:sameAs/*
dc:Event/Loc:aut/*/rda:dateOfBirth/*/owl:sameAs/*
dc:Event/Loc:aut/*/rda:dateOfDeath/*/owl:sameAs/*     
Reading the sequences is much easier with the path browser.

     }
	\label{fig:sankey}
\end{figure}

\emph{Path outlines} being composed of sequences of properties, it would be possible to represent them with a 
 Sankey
diagram~\cite{schmidt2008sankey,riehmann2005interactive}, as shown in \autoref{fig:sankey}-a. However, the number of \emph{path outlines} that could be displayed would be limited, and it would be difficult to follow the edge that the labels relate to and to identify sequences. The \emph{path browser} keeps the links, but merges the nodes, so that the links do not need to be curved any more: they become rectangles (\autoref{fig:sankey}-b). Merged nodes are turned into vertical rectangles representing entities, allowing to display their \texttt{rdf:type} when it is known. The vertical rectangles are aggregated by property, and the height of a rectangle is proportional to the frequency of the property in all paths. This allows to prioritize the readability of the best represented properties. Even in extreme cases, where the number of properties is very high, the coordination with the filter panel (\autoref{fig:steps1}-10) allows to reach a readable state very quickly: users are typically interested either in inspecting paths that are shared by most of the entities, to know which data can be queried, or in finding entities that are not well shared, in order to fix them. The panel also allows to filter on other features, and gives an overview of the available range for each feature. 

The information about sequences is made available through interactivity: hovering a property highlights all possible sequences going through it (\autoref{fig:teaser}); clicking on it selects this property, and filters out properties which are not highlighted. Selected properties form a pattern, and all \emph{path outlines} that do not match this pattern are filtered out. The search fields with autocompletion above each column also allow to form the pattern. Furthermore, patterns and statistical filters can be combined.

\subsubsection{Details-on-demand: the detail panel} 
\begin{figure}
	\centering
	\includegraphics[width=\columnwidth]{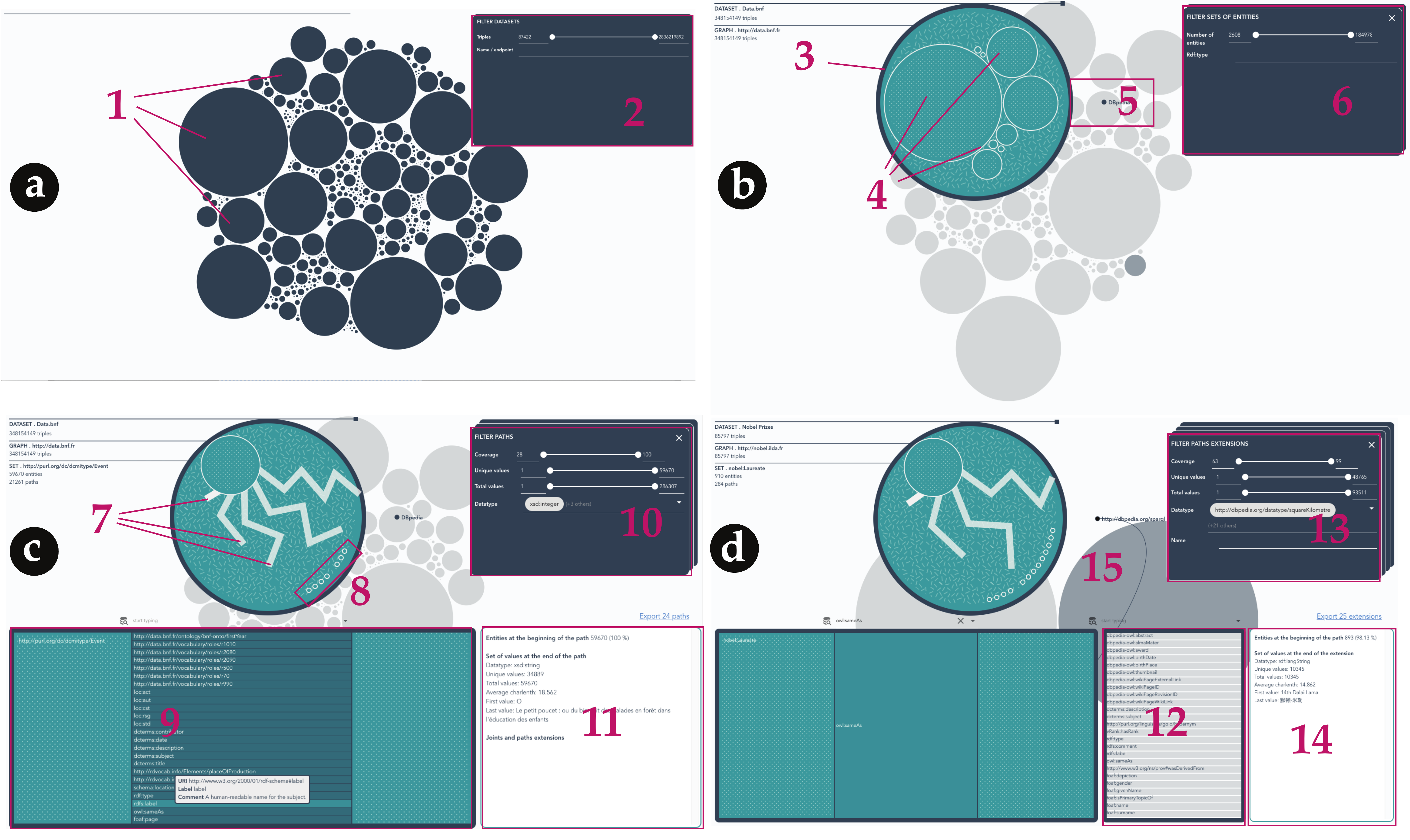}
	\caption{From overview to detail. a) At launch, the tool presents all available datasets (1), users can filter them by size and name (2). b) When a dataset is selected, interlinked datasets are placed aside (5), and sets of entities (4) are presented inside the open dataset (3). Users can filter sets of entities by size and name (6). c) When a set is selected, \emph{path outlines} of depth 1 are displayed in the \emph{Path Browser} (9), and users can select other depths (7) or other sets (8). Users can filter paths by statistical feature or name (10, 13). When a single path is hovered or selected, details are available in the detail panel (11, 14). d) When an external dataset is selected, extensions of the current path in this other dataset are presented (12) and a line points to this dataset in the context (15).
	} 
	\Description{The figure shows four screenshots. The top left one displays all available datasets with a circle packing algorithm, together with a filter panel. The top right one shows that a dataset was selected and is open in the foreground. Available sets of entities are laid out inside with a circle packing algorithm. The bottom left shows one of the sets was selected and is now on the side, with broken lines enabling to select a depth for paths. The path browser has appeared below and displays path outlines of depth one. A property is hovered, and the details for the corresponding path outlines are displayed in the detail panel, right next to the path browser. The last screenshot shows that extensions to an external dataset were selected, so the extension panel is now open between the path browser and the detail panel.
	}
	\label{fig:steps1}
\end{figure}

When users hover or selects a single \emph{path outline}, its statistical description appears in the statistical panel (\autoref{fig:steps1}-11). This panel also offers a list of linked datasets to which the selected \emph{path outline} can be extended. When a linked dataset is selected, a column is added on the right (\autoref{fig:steps1}-12), to let them browse possible extensions to the \emph{path outline}. The filter panel  (\autoref{fig:steps1}-13) and statistical panel (\autoref{fig:steps1}-14) now apply to the extended \emph{path outlines}. A line shows the target dataset, inviting users to click it and explore its \emph{path outlines}.

\subsection{Scenario of use}
\subsubsection{Scenario 1} 
A member of the \texttt{DBpedia} community would like to check the quality of the data describing music albums in the \texttt{DBpedia} dataset. She opens \PO, searches \texttt{DBpedia} in the filter panel (\autoref{fig:steps1}-a2). A dozen of datasets remain, all other are filtered out (\autoref{fig:steps1}-a1). Hovering them she can see each one corresponds to a different language. She clicks on the French version which opens in the foreground (\autoref{fig:steps1}-b3). To find music albums among the many sets of entities, she types \textit{music} in the filter panel (\autoref{fig:steps1}-b6). Five sets of entities correspond to this keyword (\autoref{fig:steps1}-b5), she hovers them and identifies \textit{schema:MusicAlbum}, which she selects. This isolates the set, displays its broken (out)lines (\autoref{fig:steps1}-c7), and opens the path browser (\autoref{fig:steps1}-c8). By default, paths of depth 1 (such as \texttt{http://dbpedia.org/ontology/composer} or \texttt{http://dbpedia.org/ontology/format}) are displayed. The interface announces that there are more than 41 000 albums, with 87 paths of depth 1. She wants to check properties with bad coverage, to see if there is a reason for this. She uses the cursor in the filter panel (\autoref{fig:steps1}-c10) to select paths with coverage lower than 10\%. She hovers available paths and inspects their coverage. She notices that the property \texttt{http://fr.dbpedia.org/property/writer} is used only once. A property which sounds very similar, \texttt{http://dbpedia.org/property/writer}, is used more than 800 times. To identify the entity she needs to modify, she clicks on the button ``See query'' that opens the SPARQL endpoint in a new window, prefilled with a query to access the set of DISTINCT values at the end of the path.
She will now do similar checks with other paths of depth 1 and paths of depth 2.

\subsubsection{Scenario 2} 
A person in charge of the \texttt{Nobel} Dataset would like to know what kind of geographical information is available for the \texttt{nobel:Laureates}. Could she draw maps of their birthplaces or affiliations?  She knows there are no geo-coordinates in the dataset, but some should be available through similarity links. She opens \PO, searches \textit{nobel} in the filter panel, and opens her dataset. She then selects the \textit{nobel:Laureates} start set. She starts to look for laureates having an affiliation aligned with another dataset. She selects paths of depth 3. In the first column, she types \textit{affiliation}. This removes other properties than \textit{nobel:affiliation} from this column, and properties which are not used in a path starting with \textit{nobel:affiliation} from other columns. Among properties remaining in the second column, she can easily identify \textit{dbpedia:city}, which she selects. In the third column, she selects \texttt{owl:sameAs} property. A single path is now selected, summary information appears in the inspector: 72\% of the laureates have an affiliation aligned with an external dataset. She selects the link to display extensions in DBpedia. A list of 78 available properties to extend the path in DBpedia appear. She types \textit{geo} in the search field. A list of 4 properties containing \textit{geo:lat} and \textit{geo:long} remains. She inspects the summary information of the extended paths: only 32\% of the laureates have geo-coordinates in DBpedia. She repeats the same operations for birthplaces: 96\% have a similarity link to an external dataset, among which 61\% have geo-coordinates in DBpedia.
She can now assess the coverage of the dataset regarding the laureates and their locations, and plan to fix the missing information. 

\subsection{Implementation}
The front-end interface is developed with NodeJS, it uses Vue.js and  d3.js frameworks. The code is open source\footnote{url of the gitlab repository, anonymised for submission}.

\section{User study 2: evaluating Path Outlines}
\md{We designed an experiment to compare \PO with the virtuoso SPARQL query editor (hereafter \SV). Although comparing a non-graphical tool with a graphical tool can be controversial, it is the relevant baseline in this case: a SPARQL editor is the only way to fully perform the tasks we are evaluating as of today, and this specific editor is the most used by our target users, as confirmed by participants in study 1. 
The experiment was a} $2 \times 2  \times 3$ within-subject controlled experiment, with a mixed design (counterbalanced for the two first variables, and ordered for the last one), to compare \PO with \SV. The first independent variable was the tool, with two modalities: Path Outlines vs \SV. The second independent variable was the dataset, with two modalities: Nobel dataset vs.\ Persée dataset. The third independent variable was the task, with 3 modalities: 3 tasks ordered by difficulty (with small adaptations to the dataset). The dependent variables we collected were the perceived comfort and easiness, the execution time, the rate of success and number of errors, and the accuracy of memorising the main features of a dataset.
Our hypotheses were:
\begin{description}
\item[H1:] \PO is easier and more comfortable to use than \SV
\item[H2:] \PO leads to shorter execution time than \SV
\item[H3:] \PO leads to better task completion and fewer errors than \SV
\item[H4:] \PO facilitates recalling the main features of a dataset compared to \SV
\end{description}

\subsection{Participants}	
We recruited 36 participants (30 men and 6 women) via calls on semantic web mailing lists and Twitter, with the requirement that they should be able to write SPARQL queries. 5 participants in the interview also registered for the experiment. Job categories included 12 researchers, 10 PhD students, 9 engineers and 3 librarians. 29 produced RDF data and 31 reused them. Their experience with SPARQL ranged from 6 months to 15 years, the average being 5.07 years and the median 4 years \footnote{SPARQL has existed since 2004, the standard was released in 2008}. 12 rated their level of comfort with SPARQL as \textit{very comfortable}, 11 as \textit{rather comfortable}, 10 as \textit{fine}, and 3 as \textit{rather uncomfortable}. 18 used it  \textit{several times a week}, 13 \textit{several times a month}, 2 \textit{several times a year} and 3 \textit{once a year or less}. \md{23 of them listed Virtuoso among the tools they were using regularly.} All participation was voluntary and without compensation.

\subsection{Setup}
The experiment was mostly supervised online through a videoconferencing system.
It was run face-to-face for 3 participants. We used an online form to guide participants through the tasks and collect the results. The form provided links to our tool, to a web interface developed in JavaScript, and to a SPARQL endpoint we had set up for the experiment. In 5 cases, due to restrictions in the network, we replaced the endpoint by the \texttt{Nobel} public endpoint. We used two datasets, Nobel and Persée, which had been analysed with our tool and are hosted in our endpoint. 2 participants stopped after 2 tasks because of personal planning reasons, so we asked the last two participants to complete only 2 tasks to keep the 4 configurations balanced for all tasks.

\subsection{Tasks}
We designed 3 real world tasks, ordered by difficulty. They involved the 3 nuclear tasks that our interface supports, combined in different ways.
On Nobel Dataset, Task 1 (T1) was: Consider all the awards in the dataset. For what percentage of them can you find the label of the birthplace of the laureate of an award?
Task 2 (T2) was: Consider all the laureates in the dataset. Find all the paths of depth 1 or 2 starting from them and leading to a piece of temporal information. Indicate the data type of the values at the end of the path.
Task 3 (T3) was:  Imagine you want to plot a map of the universities. The most precise geographical information about the universities in the dataset seems to be the cities, which are aligned to DBpedia through similarity links \entity{owl:sameAs}. Find one or several properties in DBpedia (\url{http://dbpedia.org/sparql}) that could help you place the cities on a map.
The tasks on Persée Dataset were equivalent, with small adaptations to the context.

\subsection{Procedure}	
We sent an email to the participants with a link to the video conference. As they connected, we gave them a link to the form. 
They were invited to read the consent form. 
We started with a set of questions about their experience with SPARQL. Then we introduced the experiment and explained how it would unfold.
The first task T1 was displayed, associated with a technique and a dataset. We read it aloud and rephrased the statement until it made sense to the participants. 
Participants were asked to describe their plan before they performed the task. We rated the precision: 0 for no or very imprecise planning, 1 for imprecise planning, 2 for very precise planning. The time to perform the task was limited to eight minutes. If they were not able to complete in time, they were asked to estimate how much time they think they would have needed. Then they rated the difficulty of the task and the comfort of the technique.
The next task was the equivalent task T1 associated with the other technique on the other dataset. We counterbalanced the order of the technique and dataset factors, resulting in 4 configurations. After the set of two equivalent tasks, participants were asked which environment they would choose if they had both at their disposal for such a task. 
The same was repeated for tasks T2, and then T3. 
In the end, participants answered a multiple-choice query form about the general structure of a dataset: number of triples, classes, paths of length 1 and length 2. 
To finish with, they were invited to comment on the tool and make suggestions.

\subsection{Data collection and analysis}	
We collected the answers to the form, 
and analysed with R.

\subsection{Results}
\begin{figure}
	\includegraphics[width=\columnwidth]{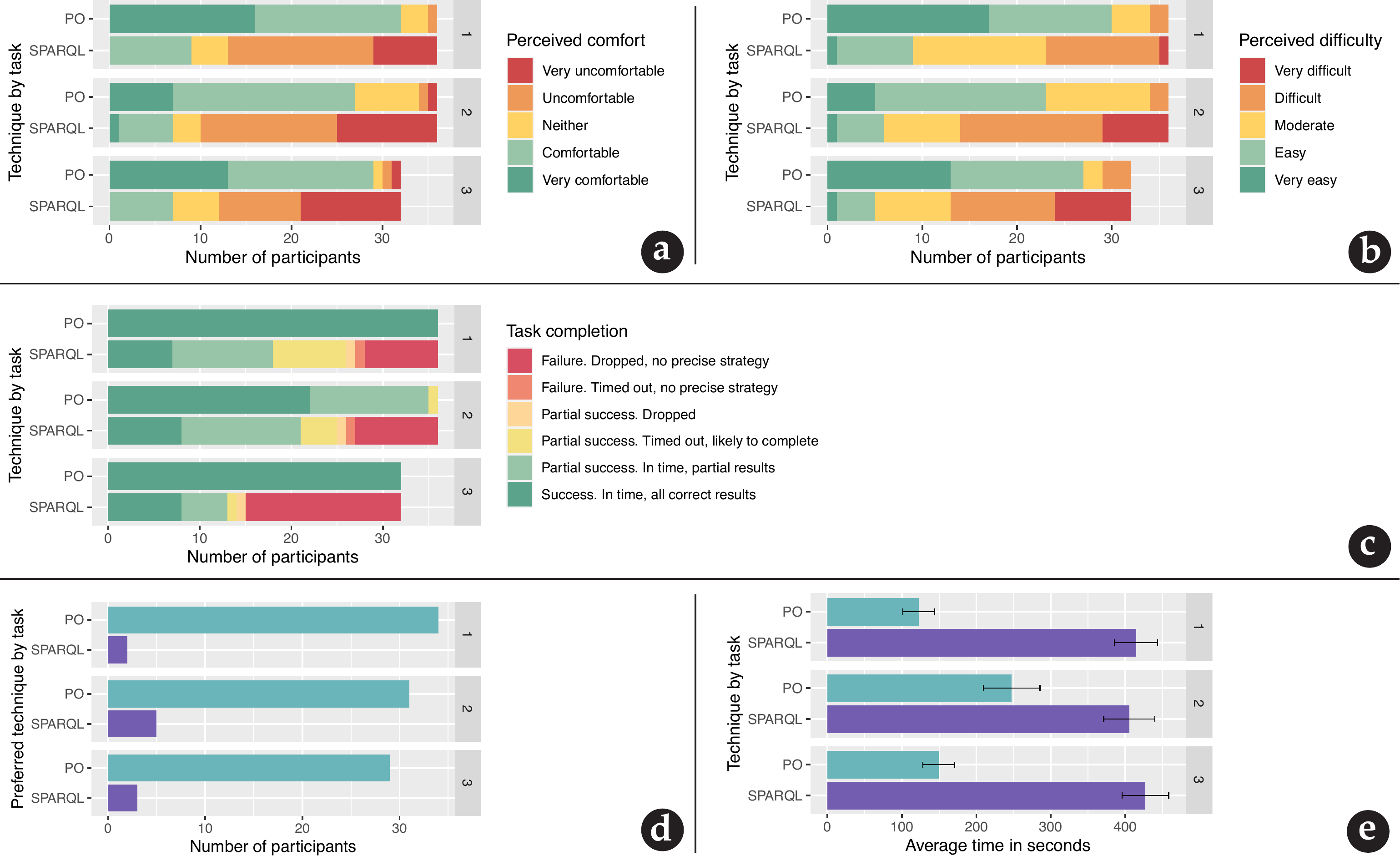}
	\caption{Comparison of \PO(PO) and \SV(SPARQL) on 3 tasks. a) and b) are on a Likert-Scale. a) Participants find \PO more comfortable, b) they perceive similar tasks as easier when performed with it, c) they are abler to complete the tasks successfully with it, d) they are quicker with it and e) prefer it to \SV.
          }
	\Description{
On the first row, the left stacked chart, labeled 'a', reports the perceived comfort reported by participant on each task with each technique. It can be converted into the following table:
	Task         Technique            Perceived comfort          Number of participants
	Task 1       Path Outlines        Very comfortable           16
	Task 1       Path Outlines        Comfortable                  16
	Task 1       Path Outlines        Neither                            3
	Task 1       Path Outlines        Uncomfortable                1
	Task 1       Path Outlines        Very uncomfortable        0
	Task 1       SPARQL-V             Very comfortable            0
	Task 1       SPARQL-V             Comfortable                   9
	Task 1       SPARQL-V             Neither                           4
	Task 1       SPARQL-V             Uncomfortable               16
	Task 1       SPARQL-V             Very uncomfortable         7
	Task 2       Path Outlines        Very comfortable             7
	Task 2       Path Outlines        Comfortable                   20
	Task 2       Path Outlines        Neither                             7
	Task 2       Path Outlines        Uncomfortable                 1
	Task 2       Path Outlines        Very uncomfortable         1
	Task 2       SPARQL-V             Very comfortable             1
	Task 2       SPARQL-V             Comfortable                    6
	Task 2       SPARQL-V             Neither                            3
	Task 2       SPARQL-V             Uncomfortable               15
	Task 2       SPARQL-V             Very uncomfortable        11
	Task 3       Path Outlines        Very comfortable            13
	Task 3       Path Outlines        Comfortable                   16
	Task 3       Path Outlines        Neither                             1
	Task 3       Path Outlines        Uncomfortable                 1
	Task 3       Path Outlines        Very uncomfortable          1
	Task 3       SPARQL-V             Very comfortable             0
	Task 3       SPARQL-V             Comfortable                    7
	Task 3       SPARQL-V             Neither                            5
	Task 3       SPARQL-V             Uncomfortable                 9
	Task 3       SPARQL-V             Very uncomfortable         11
On the first row, the right stacked chart, labeled 'b', reports the perceived difficulty reported by participant for each task with each technique. It can be converted into the following table:
	Task         Technique            Perceived difficulty          Number of participants
	Task 1       Path Outlines        Very easy                        17
	Task 1       Path Outlines        Easy                                13
	Task 1       Path Outlines        Moderate                          4
	Task 1       Path Outlines        Difficult                            2
	Task 1       Path Outlines        Very difficult                     0
	Task 1       SPARQL-V             Very easy                         1
	Task 1       SPARQL-V             Easy                                 8
	Task 1       SPARQL-V             Moderate                       14
	Task 1       SPARQL-V             Difficult                         12
	Task 1       SPARQL-V             Very difficult                    1
	Task 2       Path Outlines        Very easy                         5
	Task 2       Path Outlines        Easy                               18
	Task 2       Path Outlines        Moderate                        11
	Task 2       Path Outlines        Difficult                           2
	Task 2       Path Outlines        Very difficult                   0
	Task 2       SPARQL-V             Very easy                        1
	Task 2       SPARQL-V             Easy                                5
	Task 2       SPARQL-V             Moderate                        8
	Task 2       SPARQL-V             Difficult                         15
	Task 2       SPARQL-V             Very difficult                   7
	Task 3       Path Outlines        Very easy                      13
	Task 3       Path Outlines        Easy                              14
	Task 3       Path Outlines        Moderate                         2
	Task 3       Path Outlines        Difficult                           3
	Task 3       Path Outlines        Very difficult                   0
	Task 3       SPARQL-V             Very easy                        1
	Task 3       SPARQL-V             Easy                                4
	Task 3       SPARQL-V             Moderate                        8
	Task 3       SPARQL-V             Difficult                         11
	Task 3       SPARQL-V             Very difficult                   8

On the second row, the stacked chart, labeled 'c', reports the task completion status for each task with each technique. It can be converted into the following table:
	Task         Technique             Task completion status                                       Number of participants
	Task 1       Path Outlines        Failure. Dropped, no precise strategy                   0
	Task 1       Path Outlines        Failure. Timed out, no precise strategy                 0
	Task 1       Path Outlines        Partial success. Dropped                                       0
	Task 1       Path Outlines        Partial success. Timed out, likely to complete        0
	Task 1       Path Outlines        Partial success. In time, partial results                   0
	Task 1       Path Outlines        Partial success. In time, all correct results            36
	Task 1       SPARQL-V             Failure. Dropped, no precise strategy                    8
	Task 1       SPARQL-V             Failure. Timed out, no precise strategy                  1
	Task 1       SPARQL-V             Partial success. Dropped                                        1
	Task 1       SPARQL-V             Partial success. Timed out, likely to complete        1
	Task 1       SPARQL-V             Partial success. In time, partial results                  11
	Task 1       SPARQL-V             Partial success. In time, all correct results              7
	Task 2       Path Outlines        Failure. Dropped, no precise strategy                     0
	Task 2       Path Outlines        Failure. Timed out, no precise strategy                   0
	Task 2       Path Outlines        Partial success. Dropped                                         0
	Task 2       Path Outlines        Partial success. Timed out, likely to complete         1
	Task 2       Path Outlines        Partial success. In time, partial results                   13
	Task 2       Path Outlines        Partial success. In time, all correct results              22
	Task 2       SPARQL-V             Failure. Dropped, no precise strategy                      9
	Task 2       SPARQL-V             Failure. Timed out, no precise strategy                    1
	Task 2       SPARQL-V             Partial success. Dropped                                          1
	Task 2       SPARQL-V             Partial success. Timed out, likely to complete          4
	Task 2       SPARQL-V             Partial success. In time, partial results                    13
	Task 2       SPARQL-V             Partial success. In time, all correct results                8
	Task 3       Path Outlines        Failure. Dropped, no precise strategy                       0
	Task 3       Path Outlines        Failure. Timed out, no precise strategy                     0
	Task 3       Path Outlines        Partial success. Dropped                                          0
	Task 3       Path Outlines        Partial success. Timed out, likely to complete          0
	Task 3       Path Outlines        Partial success. In time, partial results                     0
	Task 3       Path Outlines        Partial success. In time, all correct results              32
	Task 3       SPARQL-V             Failure. Dropped, no precise strategy                     17
	Task 3       SPARQL-V             Failure. Timed out, no precise strategy                    0
	Task 3       SPARQL-V             Partial success. Dropped                                          1
	Task 3       SPARQL-V             Partial success. Timed out, likely to complete          1
	Task 3       SPARQL-V             Partial success. In time, partial results                     5
	Task 3       SPARQL-V             Partial success. In time, all correct results               8
	On the third row, the left stacked chart, labeled 'd', reports the preferred technique reported by participants on each task. It can be converted into the following table:	
	Task          Preferred technique            Number of participants
	Task 1       Path Outlines                      34
	Task 1       SPARQL-V                           2
	Task 2       Path Outlines                      31
	Task 2       SPARQL-V                           5
	Task 3       Path Outlines                      29
	Task 3       SPARQL-V                           3
	On the third row, the right stacked chart, labeled 'e', reports the average time on each task with each technique. It can be converted into the following table:
	Task         Technique            Average time              Confidence interval
	Task 1       Path Outlines       122.6944                   21.35482
	Task 1       SPARQL-V            414.3889                   29.01415
	Task 2       Path Outlines       244.0556                   39.85904
	Task 2       SPARQL-V            405.3889                   34.31149
	Task 3       Path Outlines       134.8438                    22.97322
	Task 3       SPARQL-V            427.0312                   31.39090
	}
	\label{fig:figure2}
\end{figure}

\subsubsection{Perceived comfort and easiness}
In general, participants found \PO more comfortable than \SV (\autoref{fig:figure2}a). Several participants said that they would need more time to become fully comfortable with \PO. Five minutes of practice was indeed a very short time, but the level of comfort reported with \PO is already quite satisfactory. 
The level of comfort reported when performing tasks with \SV was lower than the level initially expressed. We interpret this as being due partly to the fact that it is uncomfortable to code when an experimenter is watching, and partly to the difficulty of the tasks. Being very familiar with SPARQL does not mean being familiar with queries involving both sets of entities and deep paths. This supports the idea that a specific tool for such tasks can be useful even for experts. Three users mentioned being less comfortable with Virtuoso than with their usual environment. However, Virtuoso was the tool most frequently listed as usual by participants (23).	
Participants perceived the same tasks as being easier when performed with \PO than with \SV, as shown in \autoref{fig:figure2}b. We think this is because \PO enables them to manipulate directly the paths, saving them the mental process of reconstructing the paths by chaining statements and associating summary information to them. A participant wrote us an email after the experiment to thank us for the work, saying that \textquote{such tools are needed due to the conceptual difficulties in understanding large complex datasets}.
Those results are in agreement with H1.

\subsubsection{Task execution time}

We counted 8 minutes for each timeout or dropout. Participants were quicker with \PO on the three tasks, as shown in \autoref{fig:figure2}e, in agreement with H2. We applied paired sample t-tests to compare execution time, with a log transformation to normalize the distribution, with each technique for each task. 
There was a significant difference in the three tasks: T1: $t = 14.368, p = 3.026^{-16}$, T2: $t = 6.3173, p = 2.956^{-7}$, T3: $t = 17.467, p < 2.2^{-16}$, which shows that participants were significantly faster on each task with \PO than with \SV. The effect size is very large: the median is 480s with \SV vs.\ 119s with \PO on T1,  472.5s with \SV vs.\ 215s with \PO on T2, and 480s with \SV vs.\ 146.5s with \PO on T3. 
Those who did not complete the tasks were asked to give an estimation of the additional time they would have needed. We did not use self-estimations to make a time comparison since not all participants were able to answer, and such estimations are likely to be unreliable since time perception and self-perception are influenced by many factors. However, we report them as an indicator: for participants with a very precise plan, it ranged from 30~seconds to one hour; with an imprecise plan, it ranged from 15~seconds to 45 minutes; and with no plan, it ranged from 4 minutes to several hours. Task 2 required them to look at paths of two different depths.
Although participants were longer on this task, \PO still outperformed Virtuoso SPARQL query editor, but several participants expressed the wish to see both depths at the same time.

\subsubsection{Task completion and errors} 
Using our tool, only one participant timed out on task 2, all others managed to complete each of the tasks within 8 minutes. With \SV, there were 37 dropouts (9 on T1, 10 on T2 and 18 on T3) and 15 timeouts (9 on T1, 5 on T2 and 1 on T3). Among the tasks completed in time, 28 did had erroneous or incomplete results with \SV (11 on T1, 13 on T2 and 5 on T3) versus 13 with our tool (on T2), as summed up in \autoref{fig:figure2}c. 
The main errors on T1 were that some participants counted the number of paths matching the pattern instead of the number of documents having such paths (either by counting values at the end of the paths or by counting entities without the DISTINCT keyword). It occurred 9 times in \SV, and never with our tool. Four participants were close to making the mistake but corrected themselves with \SV, and one did so with our tool. Another error occurred only once with \SV: the participant started from the wrong class of resource.
T2 presented the particular difficulty that temporal information in RDF datasets can be typed with various data types, including \texttt{xsd:string} and \texttt{xsd:integer}. The most common error was to give only part of the results, either because of relying on only one data type, or because it was difficult to sort out the right ones when displaying all of them. It occurred 12 times with both techniques. The mean percentage of correct results was 75\% with our tool, versus 50\% with \SV. With \SV, one participant happened to give all paths as an answer, including non-temporal ones, which we regarded as a partial success. 
For T3, one participant gave an answer that did not meet the requirement with \SV, stating that it would be too complicated. Another error which happened 5 times was that the query timed out, although it was correct. There are tricks and workarounds, but in most cases, the time needed to write the query and realise it would time out was already too long to start figuring out a workaround. This is a common problem with federated queries on sets, also reported by Warren and Mulholland~\cite{warren2018using}.
Overall, our results are in agreement with H3.

\subsubsection{Memorising the main features of a dataset}
At the end of the experiment, participants answered MCQ questions about the structure of both datasets. Answers were very sparse, most participants did not remember the information at all, and there was no significant difference between the techniques. We cannot make any conclusion from the data we collected. We think this is related to the fact that participants were fully focused on finishing the tasks in time, and did not have time to look at contextual elements of the interface.
Therefore, the results are not in agreement with H4.

\subsubsection{Preference}
Most participants preferred \PO (34 on T1, 31 on T2 and 29 on T3) versus Virtuoso SPARQL query editor (2 on T1, 5 on T2 and 3 on T3), as shown in \autoref{fig:figure2}b. 

\subsubsection{Other user comments}\label{sec:other}
Several participants expressed the need for such a tool as \PO in their work and asked if they could try it on their own data. Most of them liked the tool and made positive comments. One participant wrote an email after the experiment to thank us for the work, saying that \textquote{such tools are needed due to the conceptual difficulties in understanding large complex datasets}. It is interesting to note that the participant happened to be one of the two participants who had difficulties to relate to the tasks during the interview.

\medskip

Our tasks are difficult to perform with SPARQL because they need to be decomposed in many steps, combining several types of difficulties, and 
they require to think in two dimensions: broad to consider sets of entities and objects, and deep to traverse the graph. This is not intuitive, and the cognitive load to remember the sequences of path is heavy. 
Our tool only required to browse and select, as it used the granularity required by the task. 

\section{Discussion and Conclusion}
RDF data producers face a challenge: the particular structure of their data questions the efficiency of traditional summarisation and visualisation techniques. To address this issue, we presented the concept of \emph{path outlines}, to produce path-based summaries of RDF data, with an API to analyse them. We interviewed 11 data producers and confirmed their interest. We designed and implemented \PO, a tool to support data producers in browsing path-based summaries of their datasets. We compared \PO with \SV. \PO was rated as more comfortable and easier. It performed three times faster and lowered the number of dropouts, despite the fact that participants had, on average, 5 years of experience with SPARQL versus 5 minutes with our tool.

We used coordinated views combining new visualisations with filter and detail panels to support the representation and manipulation of those complex objects. A limitation of our combination is that it relies on splitting the paths by depth. While this enabled us to display very high numbers of paths, there are cases where users would prefer to see several depths at the same time, as for Task 2. With the current interface, this means repeating the same task with different depths. In future work, we would like to investigate solutions to go from one depth to another more easily, and/or to inspect several depths at the same time. 

The concept of \textit{path outlines} can be developed to support a wider range of metrics, such as the detection of outliers suggested by a participant in the first study. To go further in this direction, integrating statistics with the content~\cite{perer2009integrating} could make the path overview the entry point for an iterative analysis of the content, as advanced profiling tools in other databases communities start to do~\cite{kandel2012profiler}. This would allow to address more elaborate tasks, such as finding the reasons for a problem pointed out by the summary.
Associated to other visualisations, that are still to design and implement, the concept  can have many applications. For instance, it can support ontologists in bettering the quality of RDF data models, showing how a modification of a property in the model would impact the potential paths traversing it, addressing  their needs to ``make changes to the inferred hierarchy explicit''~\cite{Vigo:2015:CCK:2702123.2702495}. 

We believe that the development of Knowledge Graphs will benefit from path-based summaries and tools such as \PO, presenting information to users at a granularity and form matching their needs to make sense of the information contained in Knowledge Graphs.
%
 We think that such tools will help overcome some of the complexity at the heart of Knowledge Graphs due to atomizing data as RDF triples, and will leverage high-quality Knowledge Graphs.


\bibliographystyle{ACM-Reference-Format}
\bibliography{main}


\end{document}